\documentclass[aps,prd,showpacs,nofootinbib,twocolumn]{revtex4}
\usepackage{latexsym}
\usepackage{amsmath,amsfonts}
\usepackage{amsbsy}
\usepackage{mathrsfs}
\usepackage{color}
\usepackage{float}
\usepackage{dsfont}
\usepackage{psfrag}
\usepackage{enumerate}

\usepackage{amsmath,amssymb,calc,amsfonts}
\usepackage{latexsym}
\usepackage{hyperref}

\usepackage{graphicx,calc,epsfig}

\begin{document}

\title{Weak lensing in a plasma medium and gravitational deflection of massive particles using the Gauss-Bonnet theorem. A unified treatment.
}

\date{\today}

\author{Gabriel Crisnejo$^1$ and Emanuel Gallo$^1,^2$}

\affiliation{$^1$FaMAF, UNC; Ciudad Universitaria, (5000) C\'ordoba, Argentina \\ $^2$Instituto de F\'isica Enrique Gaviola (IFEG), CONICET, \\
Ciudad Universitaria, (5000) C\'ordoba, Argentina. }

\begin{abstract}

We apply the Gauss-Bonnet theorem to the study of light rays in a plasma medium in a static and spherically symmetric gravitational field and also to the study of timelike geodesics followed for test massive particles in a spacetime with the same symmetries. The possibility to using the theorem follows from a correspondence between timelike curves followed by light rays in a plasma medium and spatial geodesics in an associated Riemannian optical metric. A similar correspondence follows for massive particles. For some examples and applications, we compute the deflection angle in weak gravitational fields for different plasma density profiles and gravitational fields.  

\end{abstract}


\maketitle


%
\section{Introduction}\label{section1}

The study of gravitational lenses has had and continues to have a tremendous impact on our understanding of the components of the Universe and the distribution of matter. From the observational point of view we can cite many examples, in which through the study of the curvature of light the content of matter of astrophysical objects can be determined (see, for example Refs.\cite{Hoekstra:2013via,2015IAUS..311...86M,Giocoli:2013tga} and references therein). 

On the other hand, studies of gravitational lens systems are also important inthe characterization of the cosmic microwave background radiation and related cosmological aspects \cite{Lewis20061,Nguyen:2017zqu,Marozzi:2016und,Peloton:2016kbw,Fabbian:2017wfp,Pratten:2016dsm,Hagstotz:2014qea,1475-7516-2015-06-050,Marozzi:2016uob,Petri:2016qya,Schaefer:2005up,Cooray:2002mj,Marozzi:2016qxl}. At the same time, from the astrophysical point of view, its study is also necessary because it allows us to predict the shape (or shadow) of a black hole or the gravitational lensing properties of more exotic compact bodies.

In general, the expressions for the deflection angle or the associated optical scalars are written in terms of derivatives of the different components of the metrics. Notwithstanding that, in Ref. \cite{PhysRevD.83.083007}, we presented an expression for the deflection angle in terms of the curvature scalars, which was generalized to the cosmological context in Ref. \cite{Boero:2016nrd} and recently by us to second order in perturbations of a flat metric\cite{PhysRevD.97.084010}. Intriguingly, this is not the only known way to write the deflection angle in terms of curvature quantities.  Recently, Gibbons and Werner introduced an elegant new way of studying gravitational lensing using the Gauss-Bonnet theorem and an associated optical metric \cite{Gibbons:2008rj}. In particular, they obtained a relation between the deflection angle, the Gaussian curvature of the associated optical metric and the topology of the manifold. 

Since then, many and varied applications have emerged. In particular, in the last years this new technique has being used in order to compute gravitational lensing quantities in a variety of spacetimes in vacuum; electro-vacuum or with different scalar fields or effective fluids.  In Refs. \cite{Jusufi:2015laa,Jusufi:2016wiz,Jusufi:2016sym,Jusufi:2017gyu,Jusufi:2017hed,Jusufi:2017vta,Jusufi:2017drg,Jusufi:2017xnr,Jusufi:2018kmk,Jusufi:2018waj,Sakalli:2018nug,Arakida:2017hrm,Ishihara:2016vdc,Ishihara:2016sfv} we find applications of the method to the study of a variety of different spacetimes with spherical symmetry, and in \cite{Werner:2012rc} the method was modified by Werner in order to allow the study of gravitational lensing in rotating and stationary spacetimes. This new version was applied to a variety of metrics in Refs. \cite{Jusufi:2017lsl,Jusufi:2017vew,Jusufi:2017mav,Jusufi:2017uhh,Jusufi:2018jof,Ono:2017pie}. 

On the other hand, one expects that compact astrophysical objects and even galaxies or clusters of galaxies will be immersed in a plasma fluid. In general, in the visible spectrum, the modification of the gravitational lensing quantities due to the presence of the plasma is negligible, because they are only significant in the radio-wave regime. Nevertheless there exists in actuality some radio-telescope projects that work in the frequency bandwidth in which these effects could be observable \cite{1991CuSc...60..106S,2013AA...556A...2V,2009IEEEP..97.1497L,Budianu-2015,Bentum:2016ekl}. Motivated by that, a proliferation of works dealing with the influence of plasma media on the trajectory of light rays in a external gravitational field associated to compact bodies have resurged\cite{BisnovatyiKogan:2008yg,BisnovatyiKogan:2010ar,Tsupko:2013cqa,Tsupko:2014sca,Tsupko:2014lta,Perlick:2015vta,Bisnovatyi-Kogan:2015dxa,Perlick:2017fio,Bisnovatyi-Kogan:2017kii, 2013ApSS.346..513M,2016ApSS.361..226A, 2017IJMPD..2650051A,
2017IJMPD..2641011A,2017PhRvD..96h4017A,2018arXiv180203293T,Rogers:2015dla,Rogers:2016xcc,Rogers:2017ofq,2018MNRAS.475..867E}. 

For all these reasons, it would be desirable to find new ways to study this problem in situations in which there exists a plasma environment.
In this paper, inspired in the powerful Gibbons-Werner method, we use an appropriately chosen two-dimensional optical metric to extend the use of the Gauss-Bonnet theorem to spherically symmetric spacetimes in the presence of a cold nonmagnetized plasma. Even when we do not discuss nondispersive media, a similar application of the optical metric to that case is possible.  

Moreover, there exists a correspondence between the dynamics of light rays in an homogeneous plasma and massive particles following the geodesic at the same spacetime. This allows us to use the Gibbons-Werner method also in these situations. Recently, Gibbons introduced a Jacobi metric, which is basically the same we discuss here\cite{Gibbons:2015qja}. For an example, we will use this correspondence to compute the bending angle for massive particles in a Schwarzschild solution. 

This work is organized as follows. In Sec. II we briefly review the dynamic of light rays in a cold nonmagnetized plasma medium, and in particular how to use the optical metric to obtain information about the orbits of light rays through the use of the Gauss-Bonnet theorem.  In Sec. III we present a collection of known examples but also new applications for several compact objects surrounded by an homogeneous plasma. In Sec. IV, we show how this method can also be applied to the discussion of nonuniform plasma, and in particular, we show, for weak gravitational fields and small deflection angles, the equivalence between the deflection angle obtained in the framework of the Gibbons-Werner method and that which is obtained from explicit solutions of the Hamilton equations as shown by Ref. \cite{BisnovatyiKogan:2010ar}. In Sec. V, we discuss how to use this method to compute deflection angles of massive particles in spherically symmetric spacetimes.
We finalize with general comments and the prospect of future research. An Appendix with a couple of extra examples of the use of the method to nonuniform plasma medium is also included.  
\section{The optical metric and the Gauss-Bonnet theorem}\label{section2}
\subsection{The optical metric associated to a plasma medium in an external gravitational field}
Let us consider a static spacetime $(\mathcal{M},g_{\alpha\beta})$ filled with a cold nonmagnetized plasma described by the refractive index $n$ \cite{Bisnovatyi-Kogan:2015dxa,Perlick:2015vta},
\begin{equation}\label{refractive-index}
n^{2}(x,\omega(x))=1-\frac{\omega_{e}^{2}(x)}{\omega^{2}(x)},
\end{equation}
where $\omega(x)$ is the photon frequency measured by a static observer while $\omega_{e}(x)$ is the electron plasma frequency,
\begin{equation}
\omega^{2}_{e}(x)=\frac{4\pi e^{2}}{m_{e}} N(x)=K_e N(x),
\end{equation}
where $e$ and $m_{e}$ are the charge of the electron and its mass, respectively; and $N(x)$ is the number density of electrons in the plasma. Note that, only light rays with $\omega(x) > \omega_{e}(x)$ propagate through the plasma.
On the other hand, if $\omega(x) < \omega_{e}(x)$, the refractive index becomes imaginary, and the waves with such frequencies will not propagate through the plasma and will be evanescent. The reason that the plasma frequency sets the physical scale can be understood as originating in the relation between the conduction current and the displacement current. In the first place, the conduction current always opposes to the displacement current. On the other hand, if the frequency of the electromagnetic wave is bigger than the plasma frequency, then the conducting current is smaller than the displacement current and the electromagnetic propagation occurs, however, for a wave with the plasma frequency, the current density exactly cancels the displacement current, and for smaller frequencies, the conducting current becomes bigger than the displacement current, and the total effective current (conducting plus displacement) has the wrong sign to allow propagation. In the following, we will not consider this kind of situations; however, we refer to Ref.\cite{chen} for a study of propagation of electromagnetic waves in nondispersive media with a complex refractive index in curved spacetimes using an effective metric that includes absorption.

Note that, due to \eqref{refractive-index}, photons in a plasma deviate
from null geodesics of the underlying spacetime in a frequency-dependent way.  Moreover, even in the presence of an homogeneous plasma, namely with $\omega_e(x)=\text{constant}$, if the underlying spacetime produce a nontrivial gravitational redshift, that is the photon frequency $\omega$ changes along the trajectory, it produces a nontrivial dispersion through \eqref{refractive-index} and therefore allows again a deviation of the light rays from the null geodesics trajectories. Of course, this last effect is not present in a flat spacetime.

In this context, light propagation is usually described through the Hamiltonian \cite{Perlick:2015vta} (see also Ref.\cite{Perlick-book} for a complete and detailed treatment), 
\begin{equation}\label{Hamilt}
H(x,p)=\frac{1}{2}\bigg({g}^{\alpha \beta}(x)p_{\alpha}p_{\beta}+\omega_{e}^{2}(x)\bigg),
\end{equation}
where light rays are solutions of  Hamilton's equation
\begin{equation}\label{Hamil-eqn}
\ell^{\alpha}:=\frac{d x^{\alpha}}{d\tilde{s}}=\frac{\partial H}{\partial p_{\alpha}}, \ \ \frac{d p_{\alpha}}{d\tilde{s}}=-\frac{\partial H}{\partial x^{\alpha}}; 
\end{equation}
with the constraint
\begin{equation}\label{Hamit-constraint}
H(x,p)=0,
\end{equation}
and $\tilde{s}$ is an curve parameter along the light curves.

From \eqref{Hamit-constraint} we can see that, in general, light rays, instead of following timelike or null geodesics with respect to ${g}_{\alpha\beta}$,  describe timelike curves with the exception of a homogeneous plasma medium in which light rays follow timelike geodesics of $g_{\alpha\beta}$. It can be heuristically understood by noting that, even in a flat spacetime filled with a homogeneous plasma, it follows from \eqref{refractive-index}, and $n=c|{\bf{k}}|/\omega$, with $|{\bf k}|$ the norm of the wave number vector, that the dispersion relation reads $\omega^2=c^2|{\bf {k}}|^2+\omega^2_{e}$, and therefore a photon behaves as if it has an effective inertial mass $m_{\text{eff}}=\hbar \omega_e$. On the other hand in a gravitational field, and using the equivalence principle, this effective mass agrees with the gravitational mass, allowing the photon to follow timelike geodesics. These heuristic considerations were made mathematically precise by Kulsrud and Loeb, who studied electromagnetic wave packets, in Refs.\cite{Loeb}. See also Ref.\cite{B1}, in which more general dispersion relations were studied for a wider variety of plasma situations in a covariant way.

In the general case, even when the fact that they do not follow geodesics does not represent any restriction for the study of light propagation in a plasma medium, it is usually convenient to make a metric transformation under which light rays propagate as timelike geodesics (see, for example Refs.  \cite{Perlick:2017fio,Schulze-Koops:2017tkc} for the use of a conformal metric transformation for which light rays behave as timelike geodesics).

Note also that, defining the tensor
\begin{equation}
\tilde{g}^{\alpha\beta}={g}^{\alpha\beta}+(1-n^{2}(x,\omega(x))u^{\alpha}u^{\beta},
\end{equation}
the Hamiltonian \eqref{Hamilt} takes the form,
\begin{equation}
H(x,p)=\frac{1}{2}\tilde{g}^{\alpha\beta}(x,\omega(x))p_{\alpha}p_{\beta},
\end{equation}
with inverse $\tilde{g}_{\alpha\gamma}$ (defined as $\tilde{g}^{\alpha\beta}\tilde{g}_{\alpha\gamma}=\delta^\beta_\gamma$):
\begin{equation}\label{Gordon-metric}
\tilde{g}_{\alpha\beta}={g}_{\alpha\beta}+\bigg(1-\frac{1}{n^{2}(x,\omega(x))}\bigg)u_{\alpha}u_{\beta}.
\end{equation}
In all these expressions, we use the photon frequency measured by a static observer, which is at rest with respect to the plasma medium,  with normalized 4-velocity $u^{\alpha}$ with respect to ${g}_{\alpha\beta}$ given by
\begin{equation}\label{omega}
\omega(x)=-p_{\alpha}u^{\alpha},
\end{equation}
and the expression \eqref{refractive-index} for the refractive index. 

As explained in Ref. \cite{book:75670}, the tensor $\tilde{g}_{\alpha\beta}$ is not in general a metric tensor, due to its dependence on $p_\alpha$. However, for nondispersive media, it is a indeed a metric, and the light rays follow null geodesics with respect to it (see Ref.\cite{book:75670} for more details). In such situations the tensor \eqref{Gordon-metric} is known as the Gordon metric\cite{W-1923}. 

On the other hand, for the case of static spacetimes, even considering dispersive media one can use a Fermat-like principle\cite{book:75670}, in which the spatial projections of the light rays on the slices $t=\text{constant}$ which that solve Hamilton's equations \eqref{Hamil-eqn} are also spacelike geodesics of the following Riemannian optical metric:
\begin{equation}\label{optical-metric}
g_{ij}^{\text{opt}}=-\frac{n^2}{{g}_{00}}{g}_{ij}.
\end{equation}

From now on, we will restrict our attention to static  and spherically symmetric metrics surrounded by a cold nonmagnetized plasma with the same symmetries; that is, the physical spacetime is assumed to be described by a metric of the form
\begin{equation}\label{eq:phys}
{g}_{\alpha\beta} dx^{\alpha} dx^{\beta} = -A(r) dt^{2} + B(r) dr^{2} + C(r)(d\vartheta^{2}+\sin^{2}\vartheta d\varphi^{2}),
\end{equation}
and with a radial dependence of the plasma frequency, $\omega_{e}=\omega_{e}(r)$. 
Of course, we could consider a suitable coordinate system in which, instead of the three metric functions $A$, $B$ and $C$ we write the metric in terms of only two new functions; however we will retain the form \eqref{eq:phys} because we would like to write general expressions that remain valid for a large family of coordinate systems.
Note that we are neglecting the self-gravitation of the plasma. We also assume asymptotic flatness and that the plasma medium is static with respect to observers following integral curves of the timelike Killing vector field $\xi^\alpha=(\frac{\partial}{\partial t})^\alpha$. Consequently, we can take $u^{\alpha}$ as%
\begin{equation}
u^{\alpha}=\frac{\delta^{\alpha}_{t}}{\sqrt{A(r)}}.
\end{equation}
Because of the gravitational redshift, the frequency of a photon at a given radial position $r$ is given by:
\begin{equation}
\omega(r)=\frac{\omega_{\infty}}{\sqrt{A(r)}},
\end{equation}
where $\omega_{\infty}$ is the photon frequency measured by an observer at infinity. This implies that the refractive index $n$ only has a radial dependence.
Without a loss of generality we will also take $\vartheta=\pi/2$.
As we are interested in the application of the Gauss-Bonnet theorem to the determination of the bending angle,  following Gibbons and Werner\cite{Gibbons:2008rj}, we will make use of the associated two-dimensional Riemannian manifold $\left(\mathcal{M}^{\text{opt}},{g}^\text{opt}_{ij}\right)$ with optical metric \eqref{optical-metric} (restricted to the plane $\vartheta=\pi/2$), 
\begin{equation}\label{eq:opt}
d\sigma^{2}=g_{ij}^{\text{opt}} dx^{i}dx^{j}=\frac{n^{2}(r)}{A(r)} \bigg(B(r)dr^{2}+C(r)d\varphi^{2}
\bigg).
\end{equation}
This metric is conformally related to the induced metric on the spatial section $t=\text{constant}$, $\vartheta=\pi/2$, of the physical spacetime, and therefore of the physical
spacetime, and therefore it preserves the angles formed
between two curves at a given point.

\subsection{Gauss-Bonnet theorem}
Let us recall the Gauss-Bonnet theorem for a two-dimensional Riemannian manifold. The Gauss-Bonnet theorem connects the intrinsic geometry of a surface, given by the integral of the Gaussian curvature, with its topology described by the Euler characteristic number, which is a topological invariant.

\begin{figure}[H]
 \centering
\includegraphics[clip,width=78mm]{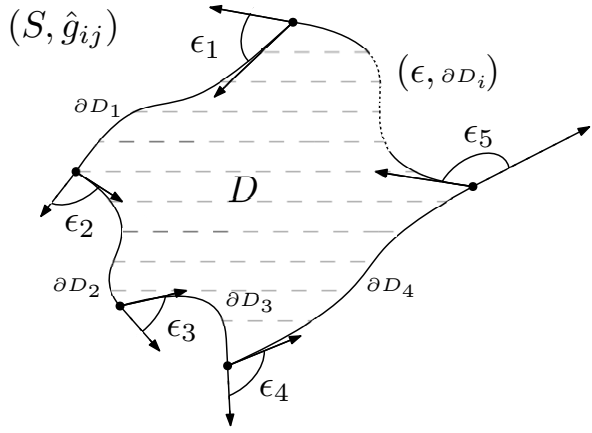}
 \caption{A region $D$ with boundary $\partial D=\cup_i \partial D_i$. In each vertex we have defined the exterior angle $\epsilon_i$ in the positive sense.}
 \label{grafico1}
\end{figure}

Precisely, this theorem can be enunciated as follows\cite{book:8292}.
Let $D\subset S$ be a regular domain of an oriented two-dimensional surface S with Riemannian metric $\hat{g}_{ij}$, the boundary of which, is formed by a closed, simple, piecewise, regular and positive oriented curve $\partial D : \mathbb{R} \supset I \to D$. Then,
\begin{equation}
\int\int_{D}\mathcal{K} dS + \int_{\partial D} \kappa_g\;d\sigma+ \sum_{i} \epsilon_{i}=2\pi\chi(D), \ \ \sigma\in I;
\end{equation}
where $\chi(D)$ and $\mathcal{K}$ are the Euler characteristic and Gaussian curvature of $D$, respectively; $\kappa_g$ is the geodesic curvature of $\partial D$ and $\epsilon_{i}$ is the exterior angle defined in the $i{\text{th}}$ vertex, in the positive sense (see Fig.~\ref{grafico1}).

Given a smooth curve $\gamma$ with tangent vector $\dot{\gamma}$ such that
\begin{equation}
\hat{g}(\dot{\gamma},\dot{\gamma})=1,
\end{equation}
and acceleration vector $\ddot{\gamma}$, the geodesic curvature $\kappa_{g}$ of $\gamma$ can be computed as,
\begin{equation}
\kappa_g=\hat{g}(\nabla_{\dot{\gamma}}\dot{\gamma},\ddot{\gamma}),
\end{equation}
which is equal to zero if and only if $\gamma$ is geodesic, because $\dot{\gamma}$ and $\ddot{\gamma}$ are orthogonal.

\begin{figure}[H]
\centering
\includegraphics[clip,width=65mm]{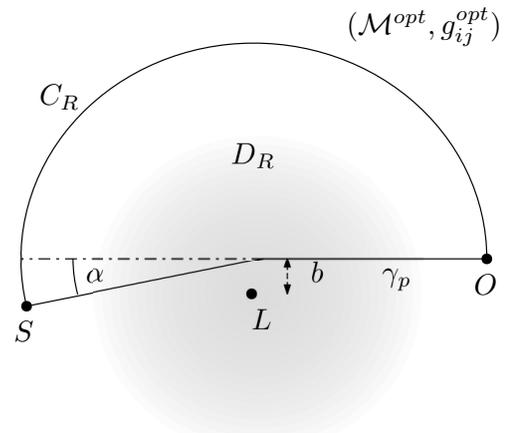}
\caption{$\partial D_{R}=C_{R}\cup\gamma_{p}$.  In this diagram, the point $S$ represents the source, and the observer is identified with $O$. $\gamma_p$ is identified with a light ray emitted by the source and  that reaches the observer at $O$. $b$ is identified with the impact parameter. The gray region represents the presence of the plasma, and $L$ represents the source of the gravitational lens. $C_R$ is a curve defined by $r(\varphi)=R=\text{constant}$. Note that all this region belongs to the two-dimensional optical manifold $\left(\mathcal{M}^{\text{opt}},g^{\text{opt}}_{ij}\right)$, and therefore the true information of the presence of the plasma medium is already codified in $g^{\text{opt}}_{ij}$.}
 \label{grafico2}
\end{figure}

Following the work of Gibbons and Werner \cite{Gibbons:2008rj}, we will apply this theorem to the optical metric $g^{\text{opt}}_{ij}$ of \eqref{eq:opt} in order to calculate the deflection angle in a plasma medium. For this, we start with the simply connected domain $D_{R}$ as shown in Fig.\ref{grafico2} with a boundary conformed by a spatial geodesic $\gamma_p$ (which codifies the information of  the light ray traveling from a source toward the observer, with an impact parameter $b$), and a curve $C_R$, defined by $r(\varphi)=R=\text{constant}$. By taking the limit of the radius $R$ of this curve going to 
infinity, and using the fact that in this limit the sum of 
the exterior angles must be equal to $\pi$ and that in the situation 
under consideration $\chi(D_{R})=1$, the resulting deflection angle $\alpha$ can be obtained from the following expression [see Fig. \eqref{grafico2} for more details]:,
\begin{equation}\label{alpha-bonnet}
\lim_{R\to\infty}\int_{0}^{\pi+\alpha}\left[\kappa_g\frac{d\sigma}{d\varphi}\right]\bigg|_{C_R}d\varphi =\pi-\lim_{R\to\infty}\int\int_{D_{R}}\mathcal{K} dS.
\end{equation}

In terms of the curvature tensor associated with the optical metric, the Gaussian curvature $K$ can be computed from
\begin{equation}\label{KR}
\mathcal{K}= \frac{R_{r\varphi r\varphi}(g^{\text{opt}})}{\text{det}(g^{\text{opt}})}.
\end{equation}
Note that in general,  it follows from \eqref{eq:opt} that 
\begin{equation}\label{dtp}
\frac{d\sigma}{d\varphi}\bigg|_{C_R}=n(R)\bigg(\frac{C(R)}{A(R)}\bigg)^{1/2}.
\end{equation}

\section{Examples and applications to homogeneous plasma medium}\label{section3}
 
To verify this approach for calculating the bending angle in the presence of a plasma medium, we will consider some illustrative examples and applications. 

Let us consider a gravitational lens surrounded by an nongravitating homogeneous plasma of which the electron number density reads
\begin{equation}
N(r)=N_{0}=\text{constant}.
\end{equation}
Note that in this case, without the presence of a nonuniform gravitational field, the refractive index should be constant, and therefore should not be a lensing effect. However, because of the presence of the gravitational field, the homogeneous plasma medium effect is not trivial. In general, this effect will be dependent of the total mass or other parameters which characterize the geometry.  
\subsubsection{Schwarzschild spacetime}
For a first example, we will calculate the bending angle for a spherically symmetric lens described by the Schwarzschild metric,
\begin{equation}\label{schw-ABC}
A(r)=1-\frac{2m}{r}, \ B(r)=\frac{1}{1-\frac{2m}{r}}, \ C(r)=r^{2},
\end{equation}
where $r>2m$. 
The refractive index for this case reads,
\begin{equation}
n(r)=\sqrt{1-\frac{\omega_{e}^{2}}{\omega^{2}_{\infty}} \bigg(1-\frac{2m}{r}\bigg)}.
\end{equation}
Thus, the associated optical metric \eqref{eq:opt} is given by,
\begin{equation}\label{opt-schw}
d\sigma^{2}=\frac{r(\omega^{2}_{\infty}-\omega_{e}^{2})+2m\omega_{e}^{2}}{(r-2m)\omega^{2}_{\infty}}  \bigg(\frac{dr^{2}}{1-\frac{2m}{r}}+r^{2}d\varphi^{2}\bigg),
\end{equation}
with determinant $\mathfrak{g}^{\text{opt}}$,
\begin{equation}\label{eq:gosch}
\mathfrak{g}^{\text{opt}}=\frac{r^{3}(\omega^{2}_{\infty}r-\omega_{e}^{2}r+2\omega_{e}^{2}m)^{2}}{(r-2m)^{3}\omega^{4}_{\infty}}.
\end{equation}
By using \eqref{KR} we get for the Gaussian curvature, 
\begin{widetext}
\begin{equation}
\mathcal{K}=\frac{\omega^{2}_{\infty}m}{r^{3}(\omega^{2}_{\infty}r-\omega_{e}^{2}r+2\omega_{e}^{2}m)^{3}} \times [ (3\omega_{e}^{2}\omega^{2}_{\infty}-2\omega^{4}_{\infty}-\omega_{e}^{4})r^{3} + (-9\omega^{2}_{\infty}\omega_{e}^{2}+3\omega^{4}_{\infty}+6\omega_{e}^{4})r^{2}m+(6\omega^{2}_{\infty}\omega_{e}^{2}-12\omega_{e}^{4})rm^{2}+8\omega_{e}^{4}m^{3}].
\end{equation}
\end{widetext}
Now, to compare the bending angle calculated using the Gauss-Bonnet theorem with expressions in the literature, we will only consider it at first order in $m$. So, we only need the following expression at linear order
\begin{equation}\label{kdssch}
\mathcal{K}dS=-\frac{2\omega^{2}_{\infty}-\omega_{e}^{2}}{r^{2}(\omega^{2}_{\infty}-\omega_{e}^{2})}m\,dr d\varphi+\mathcal{O}(m^{2}).
\end{equation}
The geodesic curvature of $C_{R}$  with respect to the metric \eqref{opt-schw} reads,
\begin{equation}
\kappa_g=\frac{\omega_{\infty}|\omega_{e}^{2}R^{2}-\omega^{2}_{\infty}R^{2}-4R\omega_{e}^{2}m+3m\omega^{2}_{\infty}R+4\omega_{e}^{2}m^{2}|}{R^{3/2}(\omega^{2}_{\infty}R-\omega_{e}^{2}R+2\omega_{e}^{2}m)^{3/2}}.
\end{equation}
On the other hand, from \eqref{dtp} and \eqref{opt-schw} it follows that for this curve,
\begin{equation}
\frac{d\sigma}{d\varphi}\bigg|_{C_R}=\frac{R}{\omega_{\infty}}\sqrt{\frac{R(\omega^{2}_{\infty}-\omega_{e}^{2})+2m\omega_{e}^{2}}{R-2m}}.
\end{equation}
Therefore, as expected for this number density profile and physical metric (which imply that the optical metric is asymptotically Euclidean) we corroborate that
\begin{equation}\label{khomons}
\lim_{R\to\infty} \kappa_g\frac{d\sigma}{d\varphi}\bigg|_{C_R}=1.
\end{equation}
At linear order in $m$, it follows using \eqref{alpha-bonnet} in the limit $R\to\infty$, and taking the geodesic curve $\gamma_p$ approximated by its flat Euclidean version parametrized as $r=b/\sin\varphi$, with $b$ representing the impact parameter in the physical spacetime that
\begin{equation}\label{alpha1}
\alpha =-\lim_{R\to\infty}\int^\pi_0\int^R_{\frac{b}{\sin\varphi}}\mathcal{K} dS.
\end{equation}
Finally, using \eqref{kdssch} the deflection angle reads 
\begin{equation}\label{phs}
\alpha=\frac{2m}{b}\bigg(1+\frac{1}{1-(\omega_{e}/\omega_{\infty})^{2}}\bigg)+\mathcal{O}(m^{2}),
\end{equation}
which agrees with the known expression found using another methods \cite{BisnovatyiKogan:2010ar}. Of course, in the absence of the plasma ($\omega_e=0$), or in the limit at which its presence is negligible ($\omega_e/\omega_\infty\to 0$)  this expression reduces to the known vacuum formula $\alpha=\frac{4m}{b}.$
\subsubsection{Schwarzschild metric pierced by a cosmic string in presence of a global monopole}
Now, we want to explore how the presence of a plasma could modify the deflection angle in the Schwarzschild metric with a global monopole characterized by a parameter $\eta$ and also pierced by a cosmic string characterized by a parameter $\mu$. This metric, was recently analyzed by Jusufi using the Gauss-Bonnet theorem in vacuum \cite{Jusufi:2015laa}. For this case, we have
\begin{equation}
A(r)=1-\frac{2m}{r}, \ B(r)=\frac{1}{1-\frac{2m}{r}}, \ C(r)=a^{2}p^{2}r^{2},
\end{equation}
where $a^{2}=1-8\pi\eta^{2}$ and $p^{2}=(1-4\mu)^{2}$ indicate the presence of a global
monopole and a cosmic string, respectively.

As the $00$ component of this metric  is the same as the Schwarzschild metric  previously considered, the photon frequency and the refractive index are the same. 

Then, the associated optical metric \eqref{eq:opt} is given by,
\begin{equation}\label{opt-schw-string}
d\sigma^{2}=\frac{r(\omega^{2}_{\infty}-\omega_{e}^{2})+2m\omega_{e}^{2}}{(r-2m)\omega^{2}_{\infty}}  \bigg(\frac{dr^{2}}{1-\frac{2m}{r}}+a^{2}\,p^{2}\,r^{2}d\varphi^{2}\bigg),
\end{equation}
with determinant $\mathfrak{g}^{\text{opt}}$,
\begin{equation}
\mathfrak{g}^{\text{opt}}=\frac{a^{2}p^{2}r^{3}(\omega^{2}_{\infty}r-\omega_{e}^{2}r+2\omega_{e}^{2}m)^{2}}{(r-2m)^{3}\omega^{4}_{\infty}}.
\end{equation}
As one expects, the Gaussian and geodesic curvatures are the same as in the Schwarzschild case.

In way simila to the previous example, we only consider $\mathcal{K}dS$ at first order in $m$, 
\begin{equation}
\mathcal{K}dS=-\frac{a\,p\, (2\omega^{2}_{\infty}-\omega_{e}^{2})}{r^{2}(\omega^{2}_{\infty}-\omega_{e}^{2})}m\,dr d\varphi+\mathcal{O}(m^{2}).
\end{equation}
On the other hand, using \eqref{dtp}, we have
\begin{equation}
\frac{d\sigma}{d\varphi}\bigg|_{C_{R}}=\frac{a\,p\,R}{\omega_{\infty}}\sqrt{\frac{R(\omega^{2}_{\infty}-\omega_{e}^{2})+2m\omega_{e}^{2}}{R-2m}},
\end{equation}
but at difference of the previous example, we obtain
\begin{equation}
\lim_{R\to\infty} \kappa_g\frac{d\sigma}{d\varphi}\bigg|_{C_R}=a\,p.
\end{equation}
At linear order in $m$, it follows from the use of \eqref{alpha-bonnet} in the limit $R\to\infty$, and taking the geodesic curve $\gamma_p$ approximated by its flat Euclidean version parametrized as $r=b/\sin\varphi$, that
\begin{equation}
\lim_{R\to\infty} \int^{\pi+\alpha}_0 \left[\kappa_g\frac{d\sigma}{d\varphi}\right]\bigg|_{C_R}d\varphi =\pi-\lim_{R\to\infty}\int^\pi_0\int^R_{\frac{b}{\sin\varphi}}\mathcal{K} dS.
\end{equation}
Hence, the deflection angle reads
\begin{equation}\label{eq:perf}
\alpha=\bigg(\frac{\pi}{a\,p}-\pi \bigg)+\frac{2m}{b}\bigg(1+\frac{1}{1-(\omega_{e}/\omega_{\infty})^{2}}\bigg)+\mathcal{O}(m^{2}).
\end{equation}
This expression generalizes the result in Ref.\cite{Jusufi:2015laa} to the case of light rays propagating in a homogeneous plasma. In particula.r if we neglect the plasma effects, $\omega_e/\omega_\infty\to 0$, Eq.\eqref{eq:perf} reduces to the expression of that reference.
\subsubsection{Self-dual lorentzian spacetimes}
In \cite{Dadhich:2001fu} a family of metrics which contains as a particular case the Schwarzschild solution was presented. This family also contain a variety of different kind of compact bodies as black holes, wormholes and naked singular geometries. 
The metric depends on three parameters $\nu$, $\lambda$ and $m$,
\begin{equation}
A(r)= \left(\nu+\lambda\sqrt{1-\frac{2m}{r}}\right)^2, \ B(r)=\frac{1}{1-\frac{2m}{r}}, \ C(r)=r^{2}.
\end{equation}
Note that in this case $\omega(r)=\omega_{\infty}\frac{\sqrt{A(\infty)}}{\sqrt{A(r)}}$ with $A(\infty)=(\nu+\lambda)^2$. Therefore, for $\nu \ne -\lambda$ (which is not asymptotically flat and will be not discussed here), we have
\begin{equation}
n(r)=\sqrt{1-\frac{\omega^2_e}{\omega^2_{\infty}}\frac{\left(\nu+\lambda\sqrt{1-\frac{2m}{r}}\right)^2}{(\nu+\lambda)^2}}.
\end{equation}
At linear order in $m$ the Gaussian curvature reads:
\begin{equation}
\mathcal{K}=-\frac{(\nu+\lambda)[(\nu+2\lambda)\omega^2_{\infty}-(\nu+\lambda)\omega^2_e]m}{(\omega^2_{\infty}-\omega^2_e)r^3};
\end{equation}
on the other hand, for the two-form $KdS$, we get
\begin{equation}
\mathcal{K}dS=-\frac{\frac{\nu+2\lambda}{\nu+\lambda}\omega^2_{\infty}-\omega^2_e}{\omega^2_{\infty}-\omega^2_e}\frac{m}{r^2} dr d\varphi.
\end{equation}
The exact expression for the geodesic curvature of $C_R$ is very cumbersome; however, at linear order in $m$ the behavior of $\kappa_g\frac{d\sigma}{d\varphi}$ for large $R$ is (as should be expected)
\begin{equation}
\kappa_g\frac{d\sigma}{d\varphi}\bigg|_{C_R}=1-\frac{m}{R}\frac{\frac{\nu+2\lambda}{\nu+\lambda}\omega^2_{\infty}-\omega^2_e}{\omega^2_{\infty}-\omega^2_e}+\mathcal{O}(R^{-2}).
\end{equation}
Finally, using the same arguments that allowed us to arrive at Eq.\eqref{alpha1}, the deflection angle reads,
\begin{equation}
\alpha=\frac{2m}{b}\frac{\frac{\nu+2\lambda}{\nu+\lambda}\omega^2_{\infty}-\omega^2_e}{\omega^2_{\infty}-\omega^2_e}.
\end{equation}
In the absence of the plasma, this expression reduces to $\alpha=\frac{2m}{b}\frac{\nu+2\lambda}{\nu+\lambda}$, which can be checked to agree with the expression obtained using alternative methods, as, for example, by solving the null geodesic equation.

For the choice of the parameters of $\nu=0$, $\lambda=1$, we recover the result for a Schwarzschild metric, Eq.\eqref{phs}.

If $\nu=\nu_0=\text{constant}$ and $\lambda=0$, which describes the so-called spatial Schwarzschild wormhole, the deflection angle  results in $\alpha=2\frac{m}{b}$; which is independent of the presence of the plasma. It is not unexpected, because in this case the refractive index is constant, due to there not being a gravitational redshift. This fact remains true for any spacetime of which the lapse function is equal to 1, as for example, for the Ellis wormhole.
 
\subsubsection{Homogeneous plasma in more general alternatives to the Schwarzschild solution}\label{subex}
Let us consider a more general class of spherically symmetric metrics with the  behavior in the components of the metric \cite{Bozza:2015haa,Kitamura:2012zy}
\begin{eqnarray}
A(r)&=&1-\frac{\mu}{r^q}+\mathcal{O}(r^{-(q+1)}),\label{a}\\
B(r)&=&1+\frac{\gamma}{r^q}+\mathcal{O}(r^{-(q+1)}),\label{b}\\
C(r)&=&r^2\bigg(1+\frac{\beta}{r^q}\bigg)+\mathcal{O}(r^{-(q-1)})\label{c},
\end{eqnarray}
with $\mu,\beta$ and $\gamma$ three parameters and $q\geq 0$.
As explained in Ref.\cite{Bozza:2015haa}, a coordinate transformation can be made such that the components of the metric preserve the form of Eqs.\eqref{a}-\eqref{c} but with $\beta=0$ (Schwarzschild-like coordinates) or $\beta=\gamma$ (isotropic coordinates). At the moment, we will keep the form \eqref{a}-\eqref{c} in order to not restrict the coordinate freedom.

This family of metrics contains as a particular case the asymptotic field limit of the Schwarzschild solution (taking $\mu=\gamma=2m$, $\beta=0$ and $q=1$) and the Ellis wormhole (taking $\mu=\gamma=0$, $\beta=a^2$ and $q=2$).
For this class of metrics, the refractive index for an homogeneous plasma reads,
\begin{equation}
n(r)=\sqrt{1-\frac{\omega^2_e}{\omega^2_{\infty}}\left(1-\frac{\mu}{r^q}\right)}.
\end{equation}
Neglecting the nonlinear terms in $\mu,\gamma$ and $\beta$; that is, terms of order $\mathcal{O}(r^{-(q+1)})$, the optical metric reads:
\begin{equation}
\begin{aligned}
d\sigma^2=&\frac{\omega^2_{\infty}[(\omega^2_{\infty}-\omega^2_e)r^q+(\gamma-\mu)\omega^2_{\infty}-\gamma\omega^2_e]}{(\omega^2_{\infty}-\omega^2_e)^2r^q}dr^2\\
&+\frac{\omega^2_{\infty}[(\omega^2_{\infty}-\omega^2_e)r^q+(\beta-\mu)\omega^2_{\infty}-\beta\omega^2_e]}{(\omega^2_{\infty}-\omega^2_e)^2r^q}r^2d\varphi^2.
\end{aligned}
\end{equation}
with determinant $\mathfrak{g}^{\text{opt}}$,
\begin{equation}
\mathfrak{g}^{\text{opt}}=\frac{(r^q+\gamma)(r^2+\beta r^{2-q})[r^q(\omega^2_\infty-\omega^2_e)+\mu\omega^2_e]^2}{r^q(r^q-\mu)^2\omega^4_\infty}.
\end{equation}
After some computations the resulting expression for $\mathcal{K}dS$ linear in $\mu$, $\beta$ and $\gamma$ reads:
\begin{equation}
\mathcal{K}dS=-\frac{[(\gamma-\beta)+q(\mu+\beta)]\omega^2_{\infty}-[(q-1)\beta+\gamma]\omega^2_e}{2(\omega^2_{\infty}-\omega^2_e)r^{q+1}}drd\varphi.
\end{equation}
The asymptotic expression for $\kappa_g\frac{d\sigma}{d\varphi}$ is
\begin{equation}
\kappa_g\frac{d\sigma}{d\varphi}\bigg|_{C_R}=1-\frac{q\omega^2_{\infty}\mu}{(\omega^2_{\infty}-\omega^2_e)R^q}-\frac{\gamma}{2R^q}+\frac{(1-q)\beta}{2R^q}+\mathcal{O}(\frac{1}{R^{q+1}}).
\end{equation}
Therefore, we can use again the Eq.\eqref{alpha1}, and the final result for the deflection angle is:
\begin{widetext}
\begin{equation}\label{eq:altang}
\alpha=\frac{\sqrt{\pi}\Gamma(\frac{q}{2}+\frac{1}{2})}{2b^q\Gamma(1+\frac{q}{2})}\frac{[(q-1)\beta+\gamma+q\mu]\omega^2_{\infty}-[(q-1)\beta+\gamma]\omega^2_e}{\omega^2_{\infty}-\omega^2_e},
\end{equation}
\end{widetext}
where 
\begin{equation}\label{Gamma-function}
\Gamma(u)=\int_{0}^{\infty}v^{u-1}e^{-v}dv,
\end{equation}
is the Gamma function.
This expression generalizes some known particular formulas considered in the literature without the presence of a plasma medium. 

In isotropic coordinates ($\beta=\gamma$) the expression \eqref{eq:altang} reduces to
\begin{equation}\label{eq:altangiso}
\alpha=\frac{\sqrt{\pi}\Gamma(\frac{q}{2}+\frac{1}{2})q}{2b^q\Gamma(1+\frac{q}{2})}\frac{(\gamma+\mu)\omega^2_{\infty}-\gamma\omega^2_e}{\omega^2_{\infty}-\omega^2_e};
\end{equation}
and in Schwarzschild-like coordinates $(\beta=0)$, it reduces to
\begin{equation}\label{eq:altangsch}
\alpha=\frac{\sqrt{\pi}\Gamma(\frac{q}{2}+\frac{1}{2})}{2b^q\Gamma(1+\frac{q}{2})}\frac{(\gamma+q\mu)\omega^2_{\infty}-\gamma\omega^2_e}{\omega^2_{\infty}-\omega^2_e}.
\end{equation}
In the absence of plasma, these expressions agree with the relations found in Refs. \cite{Bozza:2015haa,Kitamura:2012zy}. In particular, for the choice of the parameters $\mu=\gamma=2m$,  and $q=1$, Eqs.\eqref{eq:altangiso} and \eqref{eq:altangsch} reproduce the result of the deflection angle for the Schwarzschild solution. For the choice  $\mu=\gamma=0$, $\beta=a^2$, and $q=2$, using \eqref{eq:altang} we find that $\alpha=\frac{\pi a^2}{4b^2}$, which is the well-known value of the deflection angle for the Ellis spacetime at lower order in $a^2$ in the weak field approximation.

Let us focus now on the expression \eqref{eq:altangiso} for the bending angle in isotropic coordinates. From this expression, and due to the spherical symmetry, there exist simple relations that allow us to compute the useful optical quantities in weak gravitational lensing \cite{PhysRevD.83.083007} in terms of the expression for the deflection angle, namely the shear $\tilde\gamma=-\tilde\gamma(b)e^{2i\theta}$ (with $\theta$ a polar angle defined in the celestial sphere of the observer \cite{PhysRevD.83.083007}) and the convergence $\tilde\kappa$. We find in the case under study that 
\begin{eqnarray}
\tilde\gamma(b)&=&-\Delta\frac{\sqrt{\pi}\Gamma(\frac{q+3}{2})[(\gamma+\mu)\omega^2_\infty-\gamma\omega^2_e]}{b^{q+1}\Gamma(\frac{q}{2})(\omega^2_\infty-\omega^2_e)},\label{shear}\\
\tilde{\kappa}(b)&=&-\Delta\frac{\sqrt{\pi}(q-1)\Gamma(\frac{q+1}{2})[(\gamma+\mu)\omega^2_\infty-\gamma\omega^2_e]}{2b^{q+1}\Gamma(\frac{q}{2})(\omega^2_\infty-\omega^2_e)}\label{conv};
\end{eqnarray}
with $\Delta=\frac{d_ld_{ls}}{d_s}$, a scale factor dependent on the distances $d_l$, $d_{ls}$, and $d_s$ representing the distance lens observer, lens source and observer source respectively. 
Now we consider some interested families as classified in Refs.\cite{Bozza:2015haa}. Note that there exists an intersection between them. In the following, we omit writting the global factor $\Delta$. \\

\underline{\it{Family I: Extended dust distributions:}}\\

This case is obtained by requiring that $\mu=\gamma$. In such situation, the metrics coming from Eqs.\eqref{a}, \eqref{b} and \eqref{c}, can be interpreted in the framework of the Einstein equations as coming from an effective energy-momentum tensor of a perfect fluid with vanishing pressure \cite{Bozza:2015haa}.
For this family, the deflection angle, the shear and the convergence reduce to
\begin{eqnarray}
\alpha&=&\frac{\sqrt{\pi}\Gamma(\frac{q}{2}+\frac{1}{2})q\mu}{2b^q\Gamma(1+\frac{q}{2})}\frac{2\omega^2_{\infty}-\omega^2_e}{\omega^2_{\infty}-\omega^2_e},\label{angle2}\\
\tilde\gamma(b)&=&-\frac{\sqrt{\pi}\Gamma(\frac{q+3}{2})\mu}{b^{q+1}\Gamma(\frac{q}{2})}\frac{2\omega^2_\infty-\omega^2_e}{\omega^2_\infty-\omega^2_e},\label{shear2}\\
\tilde{\kappa}(b)&=&\frac{\sqrt{\pi}(1-q)\Gamma(\frac{q+1}{2})\mu}{2b^{q+1}\Gamma(\frac{q}{2})}\frac{2\omega^2_\infty-\omega^2_e}{\omega^2_\infty-\omega^2_e}\label{conv2}.
\end{eqnarray}
The Schwarzschild solution is a particular example of this family, with $q=1$ and $\mu=2m$. For this value of $q$, the converge is zero, independent of the presence of the plasma. As explained by Bozza and Postiglione \cite{Bozza:2015haa}, if $0<q<1$ and $\mu>0$, the energy-density is positive. However, if $q>1$, the convergence is negative, and it is produced by an exotic lens with an effective energy-momentum tensor with a negative mass density. In all this family, the correction factor due to the presence of the homogeneous plasma is the same as in the Schwarzschild spacetime. \\

\underline{\it{Family II: Pure anisotropic pressure distribution:}}\\

This family is characterized by $q=1$, and it results in an effective energy-momentum tensor which has zero energy density and an anisotropic pressure.
The optical scalars are:
\begin{eqnarray}
\alpha&=&\frac{(\gamma+\mu)\omega^2_{\infty}-\gamma\omega^2_e}{b(\omega^2_{\infty}-\omega^2_e)}\label{a3},\\
\tilde\gamma(b)&=&-\frac{(\gamma+\mu)\omega^2_{\infty}-\gamma\omega^2_e}{b^2(\omega^2_{\infty}-\omega^2_e)},\label{shear3}\\
\tilde{\kappa}(b)&=&0\label{conv3}.
\end{eqnarray}
If $\gamma=\mu=2m$, we recover again the Schwarzschild solution. If $\mu=0$, the plasma does not influence to the total deflection angle. The particular case $\mu=-\gamma$ will be analyzed in the case $V$.\\

\underline{\it{Family III: Constant lapse family:}}\\

This family is characterized by $\mu=0$, and therefore as there is not gravitational redshift, a homogeneous plasma does not influence in the optical scalars. Their expressions read 

\begin{eqnarray}
\alpha&=&\frac{\sqrt{\pi}\Gamma(\frac{q}{2}+\frac{1}{2})q\gamma}{2b^q\Gamma(1+\frac{q}{2})}\label{a5},\\
\tilde\gamma(b)&=&-\frac{\sqrt{\pi}\Gamma(\frac{q+3}{2})\gamma}{b^{q+1}\Gamma(\frac{q}{2})},\label{shear5}\\
\tilde{\kappa}(b)&=&-\frac{\sqrt{\pi}(q-1)\Gamma(\frac{q+1}{2})\gamma}{2b^{q+1}\Gamma(\frac{q}{2})}\label{conv5}.
\end{eqnarray}
\underline{\it{Family IV: Zero lensing family
(in the absence of plasma):}}\\

This family is characterized by $\mu=-\gamma$. Without the presence of a plasma, the total deflection angle is zero, however, in the presence of a plasma, even when it is homogeneous, the deflection angle takes a non zero value. It is due to the fact that there exists a nontrivial redshift that makes the refractive index dependent of the radial coordinate $r$, however, the spatial components of the metric cannot cancel this new contribution. The optical scalars for this situation are

\begin{eqnarray}
\alpha&=&\frac{\sqrt{\pi}\Gamma(\frac{q}{2}+\frac{1}{2})}{2b^q\Gamma(1+\frac{q}{2})}\frac{q\mu\omega^2_e}{\omega^2_{\infty}-\omega^2_e}\label{a6},\\
\tilde\gamma(b)&=&-\frac{\sqrt{\pi}\Gamma(\frac{q+3}{2})\mu\omega^2_e}{b^{q+1}\Gamma(\frac{q}{2})(\omega^2_\infty-\omega^2_e)},\label{shear6}\\
\tilde{\kappa}(b)&=&\frac{\sqrt{\pi}(1-q)\Gamma(\frac{q+1}{2})\mu\omega^2_e}{2b^{q+1}\Gamma(\frac{q}{2})(\omega^2_\infty-\omega^2_e)}\label{conv6}.
\end{eqnarray}

\underline{\it{Family V: Zero spatial curvature:}}\\

This family is characterized by $\gamma=0$. which makes the
$t=\text{constant}$ slices be flat. Curiously, for this family, the optical scalars take the same form as in the zero lensing family.\\

\begin{figure}[H] 
 \centering
 \includegraphics[clip,width=92mm]{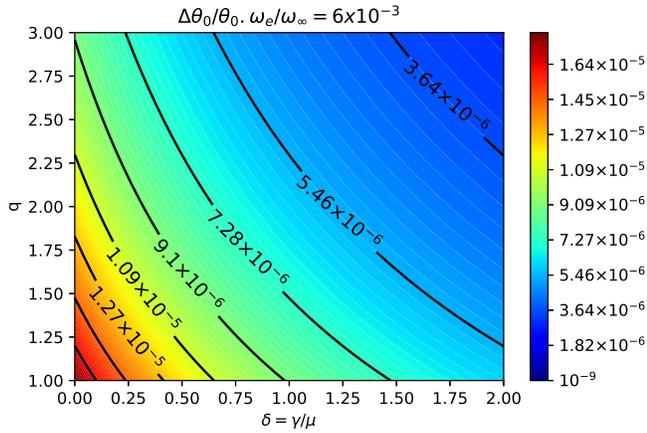}
 \caption{ Level curves of the relative change in the position of the Einstein rings $\Delta\theta_{0}/\theta_0$ for the class of metrics given by \eqref{a}-\eqref{c}. As expected, for a fixed value of the quotient $\delta=\gamma/\mu$, the relative difference becomes smaller as $q$ grows. The level curve that takes the value $9\times 10^{-6}$ (not shown) contains the particular point ($q=1,\delta=1$), corresponding to the relative position change in the Einstein rings associated with a Schwarzschild metric for the mentioned rate of frequencies.}
  \label{grafico3}
 \end{figure}

Another relevant quantity that in general changes in the presence of a plasma is the angular position of the Einstein ring. Let us assume that the parameters $\mu$ and $\gamma$ are positive. Therefore from \eqref{eq:altangiso} and the weak lens equation it follows that the
Einstein ring $\theta_{pl}$ in the presence of a plasma is given by:
\begin{equation}
\theta_{pl} = \bigg(\frac{\sqrt{\pi}\Gamma(\frac{q}{2}+\frac{1}{2})q}{2\Gamma(1+\frac{q}{2})}\frac{(\gamma+\mu)\omega^2_{\infty}-\gamma\omega^2_e}{\omega^2_{\infty}-\omega^2_e} \frac{d_{ls}}{d_{s} d_{l}^{q}}\bigg)^{\frac{1}{q+1}}.
\end{equation}
The relative change between the Einstein ring $\theta_{pl}$ in the presence of the plasma and its value $\theta_0$ in its absence is given by:
\begin{equation}\label{eq:Thetadel}
\frac{\Delta \theta_0}{\theta_0}=\frac{\theta_{pl}-\theta_0}{\theta_0}=
\bigg(\frac{1-\frac{\gamma}{\gamma+\mu}\frac{\omega^2_e}{\omega^2_\infty}}{1-\frac{\omega^2_e}{\omega^2_\infty}}\bigg)^{\frac{1}{q+1}}-1.
\end{equation}
Under the assumption that $\omega^2_e/\omega^2_\infty\ll 1$ we can approximate the above expression to 
\begin{equation}
\frac{\Delta \theta_0}{\theta_0}\approx\frac{\mu}{(1+q)(\mu+\gamma)}
\frac{\omega^2_e}{\omega^2_\infty}. \label{dtap}
\end{equation}
For an example of the magnitude of the change in the position of the Einstein ring in Fig.\eqref{grafico3}, assuming that $\mu\neq 0$ and $\delta=\gamma/\mu$ are positive, we have plotted the level curves of \eqref{dtap} for the quotient $\omega_e/\omega_\infty=6\times 10^{-3}$. For this particular frequency relation, and with a value of $\theta_0\approx 1$sec, the change is of the order of $1-10\mu sec$.  The change in the position for a Schwarzschild metric is a particular case of the level curve defined by $\delta=1$ and it was analyzed by the same value of $\omega_e/\omega_\infty$ in Ref.\cite{BisnovatyiKogan:2010ar}.

\section{Nonuniform plasma medium}\label{section4}
Now, to show how this approach to calculate the bending angle is also useful for light rays propagating in nonuniform plasma, we will recover a general expression for the deflection angle in the weak lensing regime obtained for the first time by Bisnovatyi-Kogan and Tsupko \cite{BisnovatyiKogan:2010ar} in solving the Hamilton equations. In Appendix A, we also show some explicit examples.

Let us consider an asymptotically flat and spherical symmetric gravitational lens surrounded by an inhomogeneous plasma of wich the electron number density $N(r)$ is a decreasing function of the radial coordinate $r$ and such that its radial derivative $N'(r)$ is also decreasing and smaller than $N(r)$ . In isotropic coordinates, the components of the metric in the physical spacetime is codified in the following expressions;
\begin{equation}
A(r)=1-\mu h_{00}(r), \ B(r)=1+\gamma h_{rr}(r), \ C(r)=r^2 B(r).
\end{equation}
The refractive index reads
\begin{equation}
n(r)=\sqrt{1-\frac{\omega_{e}^{2} (1-\mu h_{00}(r)) }{\omega_{\infty}^{2}}}.
\end{equation}
The associated optical metric is given by
\begin{equation}\label{h00-inhom}
d\sigma^{2}=\bigg(\frac{(1+\gamma h_{rr})(\omega_{\infty}^{2}-\omega_{e}^{2}+\mu \omega_{e}^{2} h_{00})}{\omega_{\infty}^{2}(1-\mu h_{00})}\bigg)(dr^{2}+r^{2}d\varphi^{2}),
\end{equation}
with determinant $\mathfrak{g}^{\text{opt}}$,
\begin{equation}
\mathfrak{g}^{\text{opt}}=\bigg(\frac{(1+\gamma h_{rr})(\omega_{\infty}^{2}-\omega_{e}^{2}+\mu \omega_{e}^{2} h_{00})}{\omega_{\infty}^{2}(1-\mu h_{00})}\bigg)^{2} r^{2}.
\end{equation} 
As we are only interested in terms that are linear in $\gamma$ and $\mu$, we write the Gaussian curvature at linear order in these parameters, arriving at an expression of the form
\begin{equation}                        
\mathcal{K}=\mathcal{K}_{pl}+\mu\mathcal{K}_{\mu}+\gamma\mathcal{K}_{\gamma},
\end{equation}
with
\begin{widetext}
\begin{eqnarray}
\mathcal{K}_{pl}&=&\frac{\omega_{\infty}^{2}}{2r(\omega_{\infty}^{2}-\omega_{e}^{2})^{3}} \times \bigg[K_{e}(rN{'}){'} (\omega_{\infty}^{2}-\omega_{e}^{2}) + r K_{e}^{2}N{'}^{2}\bigg],\\
\mathcal{K}_{\mu}&=&-\frac{\omega_{\infty}^{4}}{2r(\omega_{\infty}^{2}-\omega_{e}^{2})^{2}} \times \bigg[(r h_{00}{'}){'} + \mathcal{F}_{\mu}(h_{00}N{'},h_{00}N{''},h_{00}{'}N{'})\bigg], \\
\mathcal{K}_{\gamma}&=&-\frac{\omega^{2}}{2r(\omega_{\infty}^{2}-\omega_{e}^{2})} \times \bigg[(rh_{rr}{'}){'} + \mathcal{F}_{\gamma}(h_{rr}N{'},h_{rr}N{''},h_{rr}{'}N{'})\bigg],
\end{eqnarray}
\end{widetext}
and where the functions $\mathcal{F}_\mu$ and $\mathcal{F}_\gamma$ are defined as,
\begin{widetext}
\begin{eqnarray}
\mathcal{F}_{\mu}(h_{00}N{'},h_{00}N{''},h_{00}{'}N{'}) &=&  2 K_{e} (\omega_{\infty}^{2}-\omega_{e}^{2}) [h_{00} (rN{'}){'} + r N{'} h_{00}{'}] + 3 h_{00}rK_{e}^{2} N{'}^{2}, \\
\mathcal{F}_{\gamma}(h_{rr}N{'},h_{rr}N{''},h_{rr}{'}N{'}) &=& K_{e}h_{rr}[r(\omega_{\infty}^{2}-\omega_{e}^{2})N{''}+N{'}(\omega_{\infty}^{2}-\omega_{e}^{2}+K_{e}r N{'})].
\end{eqnarray}
\end{widetext}
Here, $\mathcal{K}_{pl}$ represent the plasma contribution to the Gaussian curvature which is also present in the case of a flat spacetime. $\mathcal{K}_{\mu}$ and $\mathcal{K}_{\gamma}$ take into account the deviation of the metric from the flat background, and therefore they contain not only information about the pure gravitational fields but also about the interaction between this field and the plasma.  

In principle, we could use this expression for the Gaussian curvature in order to compute the deflection angle. However, as in general the change in the deflection angle due to the presence of the refractive index is smaller than the main part due to the pure gravitational effect, we will assume as in Ref. \cite{BisnovatyiKogan:2010ar} that the deflection angle is small and therefore as a first approximation the geodesic $\gamma_p$ can be taken as the straight line geodesic of the flat Euclidean space. On the other hand, we neglect all higher-order terms of the form $\mathcal{O}(N'^2,\mu N',\mu N'',\gamma N'^2,\gamma N'')$.
Therefore in the following we discard the last term in $\mathcal{K}_{pl}$ and the terms $\mathcal{F}_\mu$ and $\mathcal{F}_\gamma$ in the other components of the Gaussian curvature. 
                      
Working at the mentioned order, we obtain for $\mathcal{K}dS$:
\begin{equation}
\mathcal{K}dS= \frac{1}{2}\bigg[\frac{K_{e}(rN{'}){'}}{\omega_{\infty}^{2}-\omega_{e}^{2}}  - \frac{\omega_{\infty}^{2}(rh_{00}{'}){'} }{\omega_{\infty}^{2}-\omega_{e}^{2}} \mu - (r h_{rr}{'}){'} \gamma\bigg]dr d\varphi.
\end{equation}
Furthermore, we need to compute $k_{g}$ and $\frac{d\sigma}{d\varphi}$ along the curve $C_{R}$ associated to the optical metric \eqref{h00-inhom}, which gives
\begin{equation}\label{in-kg}
\begin{aligned}       
\kappa_g=&\frac{\omega_{\infty}}{2R^{2}(\omega_{\infty}^{2}-\omega_{e}^{2}+\omega_{e}^{2} \mu h_{00})^{3}(1+\gamma h_{rr})^{3}(1-\mu h_{00})} \\ &\times \bigg| R K_{e}(1-\mu h_{00})^{2} (1+\gamma h_{rr}) N{'}-(1-\mu h_{00}) \\
&(\omega_{\infty}^{2}-\omega_{e}^{2}+\omega_{e}^{2}\mu h_{00}) 
(1+\gamma (r h_{rr}){'}-R \mu h_{00}{'} \omega_{\infty}^{2} \bigg|.
\end{aligned}
\end{equation}
On the other hand, we have that,
\begin{equation}\label{in-dtdphi}
\frac{d\sigma}{d\varphi}\bigg|_{C_{R}}=\frac{R}{\omega_{\infty}}\sqrt{\frac{(1+\gamma h_{rr})(\omega_{\infty}^{2}-\omega_{e}^{2}+\omega_{e}^{2}\mu h_{00})}{1-\mu h_{00}}},
\end{equation}
where all the functions in \eqref{in-kg} and \eqref{in-dtdphi} are evaluated in $r=R$.
Hence, we can check that, due to the asymptotic behavior of $h_{00}$, $h_{rr}$ and $N(r)$,
\begin{equation}
\lim_{R\to\infty} \kappa_{g}\frac{d\sigma}{d\varphi}\bigg|_{C_R}=1.
\end{equation}
Collecting all these results together we find that the deflection angle in this approximation is given by
\begin{widetext}
\begin{equation}\label{alpha-inhomo-GEN}
\begin{aligned}
\alpha&\approx -\lim_{R\to\infty}\int\int_{D_R} \mathcal{K}dS=-\int^\pi_0\int^\infty_{b/\sin\varphi} \frac{1}{2}\bigg[\frac{K_{e}(rN{'}){'}}{\omega_{\infty}^{2}-\omega_{e}^{2}}  - \frac{\omega_{\infty}^{2}(rh_{00}{'}){'} }{\omega_{\infty}^{2}-\omega_{e}^{2}} \mu - (r h_{rr}{'}){'} \gamma\bigg]dr d\varphi.
\end{aligned}
\end{equation}
\end{widetext}
Using integration by parts in the first two terms of the radial integral and neglecting again in the process the terms of order $\mathcal{O}(N'^2,h_{00}N')$,
we obtain:
\begin{equation}\label{angletsu}
\alpha\approx\int^\pi_0 \frac{1}{2}\bigg[\frac{K_{e}(rN{'})}{\omega_{\infty}^{2}-\omega_{e}^{2}} -  \frac{\omega_{\infty}^{2}(rh_{00}{'})}{\omega_{\infty}^{2}-\omega_{e}^{2}} \mu -(r h_{rr}{'})\gamma\bigg]\bigg|_{r=b/\sin\varphi}d\varphi.
\end{equation}
If we transform to a new coordinate $z$ related to $r$ as $z=\sqrt{r^2-b^2}$, thus satisfying $\tan\varphi=b/z$, we can write the expression \eqref{angletsu} as,
\begin{widetext}
\begin{equation}\label{angletsu-tsupko}
\alpha\approx-\int^\infty_{-\infty}\frac{b}{2r}\bigg[-\frac{K_{e}N{'}}{\omega_{\infty}^{2}-\omega_{e}^{2}} +  \frac{\omega_{\infty}^{2}h_{00}{'}}{\omega_{\infty}^{2}-\omega_{e}^{2}} \mu +(h_{rr}{'})\gamma\bigg]\bigg|_{r=\sqrt{b^2+z^2}}dz,
\end{equation}
\end{widetext}
which is in complete agreement with the expression (30) derived by Bisnovatyi-Kogan and Tsupko in Ref. \cite{BisnovatyiKogan:2010ar}.
For completeness, the deflection angle for two different electronic density profiles is calculated in Appendix \ref{appendix}. 

\section{Application of the Gauss-Bonnet theorem to gravitational deflection of massive particles}\label{section5}
\subsection{Optical metric}
Let us consider a static gravitational field.
As was already remark by several authors in the past (see Refs. \cite{BisnovatyiKogan:2010ar,Tsupko:2013cqa} and references therein), there exists a correspondence between the dynamic of light rays of frequency $\omega_\infty$ in a homogeneous cold nonmagnetized plasma (with characteristic frequency $\omega_e$) and the timelike geodesic motion of a test massive particle with mass $\mu$ and energy $E_\infty$ as measured by an asymptotic observer at the same gravitational field. In particular if we make the identification   $\omega_e\rightarrow\mu=\text{constant}$ , $\omega_\infty\rightarrow E_\infty$, it follows that we can use the Hamiltonian \eqref{Hamil-eqn} to describe the geodesic motion of massive particles. 

For the same reason, given any static spacetime of the form
\begin{equation}
g_{\alpha\beta}=-A(x^i)dt^2+g_{ij}dx^idx^j,
\end{equation} 
we can associate an optical metric \eqref{optical-metric} with each test particle of mass $\mu$ and energy $E_\infty$. Noting that the local energy $E(x^a)$ as measured by a static observer is related to $E_\infty$ by $E(x^a)=E_\infty/\sqrt{A(x^a)}$, it follows that the optical metric reads
\begin{equation}\label{eptm}
g^{\text{opt}}_{ij}=-\frac{n^2}{A(x^i)} g_{ij}=-\frac{1-\frac{\mu^2}{E^2_\infty}A(x^i)}{g_{00}}g_{ij}.
\end{equation}
This metric is implicit in the general work of Synge about geometrical optics in dispersive and nondispersive media (see the chapter XI of \cite{book:75670}) and also in the recent work of Gibbons where he reintroduced  (up to a constant factor $E^2_\infty$) the same metric under the name of the Jacobi metric\cite{Gibbons:2015qja}. We refer to the last reference for an elegant derivation and discussion of some of its properties.

Let us focus now in the geodesic motion of a massive particle of mass $\mu$ in and static and spherically symmetric spacetime. In particular we are interested in the description of the motion of the particle that leaves a source in an asymptotically flat region, reaches the lens at a minimal distance $r_0$, and follows its trip until an asymptotic observer. The particle is assumed leaving the asymptotic region with a speed $v$ as measured by an asymptotic observer and therefore with an energy
\begin{equation}
E_\infty=\frac{\mu}{\sqrt{1-v^2}}.
\end{equation}
In the same way let us assume that the particle has an angular momentum $J$
\begin{equation}
J=\frac{\mu vb}{\sqrt{1-v^2}},
\end{equation}
with $b$ the impact parameter.
It follows that the optical metric reads
\begin{equation}\label{eq:optmassive}
d\sigma^{2}=\frac{n^{2}(r)}{A(r)} \bigg(B(r)dr^{2}+C(r)d\varphi^{2}
\bigg),
\end{equation} 
with        
\begin{equation}\label{nmass}
n^2(r)=1-\frac{\mu^2}{E^2_{\infty}}A(r)=1-(1-v^2)A(r).
\end{equation}  
With all this information we can now study the spatial geodesics of the metric \eqref{eq:optmassive}. Note that what follows is general for any refractive index and not only for massive particles. In particular, the geodesic motion follows from the Lagrangian
\begin{equation}\label{eq:Lmassive}
\mathcal{L}=\frac{1}{2}\bigg[\frac{n^{2}(r)}{A(r)} \bigg(B(r)\bigg(\frac{dr}{d\sigma}\bigg)^{2}+C(r)\bigg(\frac{d\varphi}{d\sigma}\bigg)^{2}\bigg)\bigg],
\end{equation}
with the on-shell constraint:
\begin{equation}\label{constraint}
\frac{n^{2}(r)}{A(r)} \bigg[B(r)\bigg(\frac{dr}{d\sigma}\bigg)^{2}+C(r)\bigg(\frac{d\varphi}{d\sigma}\bigg)^{2}\bigg]=1.
\end{equation} 
From \eqref{eq:Lmassive} it follows that 
\begin{equation}\label{con}
\frac{n^2C}{A}\frac{d\varphi}{d\sigma}=\frac{J}{E_\infty}.
\end{equation}
We refer to Ref.\cite{Gibbons:2015qja} for a justification of the identification between the constant associated to this conserve quantity and $J/E_\infty$, where one must also take into account that the optical metric defined in \eqref{eptm} is related to the metric $ds^2$ used by Gibbons by $ds^2=E^2_\infty d\sigma^2$. 

From this relation and the expressions \eqref{constraint} and \eqref{con} it follows that  
\begin{equation}\label{constraint2}
\begin{aligned}
\bigg(\frac{dr}{d\varphi}\bigg)^{2}
=&\frac{C}{B}\bigg(\frac{E^2_\infty C n^2}{J^2 A }-1\bigg).
\end{aligned}
\end{equation}
The last expression for the orbital equation was also recently derived using the Hamiltonian approach \cite{Perlick:2015vta,Tsupko:2013cqa}.

Using the metric \eqref{eq:optmassive} with $n(r)$ given by \eqref{nmass} we can apply the Gauss-Bonnet theorem to the study of lensing for massive particles in any spherically symmetric gravitational field. Of course, if we want to compute the deflection angle using the Gauss-Bonnet theorem, we only need the flat trajectory of the particle written as usual, $r=b/\sin\varphi$; however the main motivation to explicitly write \eqref{constraint2} is that we will apply the Gibbons-Werner method to study the deflection angle of massive particles at second order in a Schwarzschild metric of mass $m$, and for such goal we need to know the orbit at first order in $m$.

\subsection{Application: Deflection angle of massive particles at second order in a Schwarzschild spacetime}
Here, we restrict our attention to a Schwarzschild spacetime of mass $m$ with $A(r)$, $B(r)$ and $C(r)$ given by \eqref{schw-ABC}.
Using the variable $u=1/r$,  Eq.\eqref{constraint2} reads
\begin{equation}\label{orbit}
\bigg(\frac{du}{d\varphi}\bigg)^2=-u^2+2mu^3+\frac{2m(1-v^2)}{b^2v^2}u+\frac{1}{b^2}.
\end{equation}
This equation reduces to the equation of a massless particle for $v=1$.
We want to find solutions of this equation describing the scattering of massive particles in the weak gravitational region, with the condition that the particle
comes from an asymptotic region, passes through the lens at a closer position at $\varphi=\pi/2$, and escapes to the asymptotic region again. 
To solve \eqref{orbit}, and following the approach of Ref.\cite{Arakida:2011ty} we assume that the solution can be expressed in powers of $m$ as
\begin{equation}
u=\frac{1}{b}\bigg(\sin(\varphi)+m u_1(\varphi)+m^2u_2(\varphi)\bigg)+\mathcal{O}(m^3).
\end{equation}
Hence, using the mentioned conditions, we find
\begin{widetext}
\begin{equation}\label{ugamma}
u(\varphi)=\frac{\sin\varphi}{b}+\frac{v^2\cos 2\varphi+v^2+2}{2b^2v^2}m+\frac{m^2}{16b^3}\bigg[\frac{(8+32v^2-3v^4)}{v^4}\sin\varphi+\frac{6(4+v^2)(\pi-2\varphi)}{v^2}\cos \varphi-3\sin 3\varphi\bigg]+\mathcal{O}(m^3).
\end{equation}
\end{widetext}
For $v=1$, this expression reduces to the known second-order solution of \eqref{orbit} for massless particles\cite{Arakida:2011ty}.

Now, we apply the Gauss-Bonnet theorem to compute the deflection angle to second order. 
From the optical metric that follows from \eqref{schw-ABC}, \eqref{eq:optmassive}, and \eqref{nmass}, we compute the associated determinant and Gaussian curvature,
\begin{widetext}
\begin{eqnarray}
\mathfrak{g^{\text{opt}}}&=&\frac{[2(1-v^2)m+v^2r]^2r^3}{(r-2m)^3},\\
\mathcal{K}&=&\frac{m[8(1-v^2)^2m^3+6rm^2(1-v^2)(2v^2-1)-3v^2(1-2v^2)mr^2-v^2(1+v^2)r^3]}{[2m(1-v^2)+v^2r]^3r^3}.
\end{eqnarray}
\end{widetext}
As we are interested in the computation at second order of the deflection angle, we need the expression for $\mathcal{K}dS$ at second order, which is given by:
\begin{equation}
\mathcal{K}dS=\bigg(-\frac{1+v^2}{v^2r^2}m-\frac{v^4+6v^2-4}{v^4r^3}m^2 \bigg)drd\varphi+\mathcal{O}(m^3).
\end{equation}
On the other hand, as was discussed for the homogeneous plasma in Schwarzschild solution, doing the correspondent identifications between frequencies and energy and mass it follows that \eqref{khomons} remains valid. Consequently, the deflection angle at second order is computed from \eqref{alpha1}
with $r_\gamma={u^{-1}_{\gamma}(\varphi)}$ and ${u_{\gamma}(\varphi)}$ given by the first two terms of \eqref{ugamma}. After doing the integrals, the final result for the deflection angle reads
\begin{eqnarray}\label{secondorder}
\alpha=\frac{2m}{b}\bigg(1+\frac{1}{v^2}\bigg)+\frac{3\pi}{4b^2}\bigg(1+\frac{4}{v^2}\bigg)m^2+\mathcal{O}(m^3).
\end{eqnarray}
This expression reduces to the known result for massless particles. For massive particles, there exist two different expressions in the literature. The first one given by Accioly and Ragusa \cite{Accioly:2002ck}, and the second by Bhadra,  Sarkar and Nandi \cite{Bhadra:2006fv}. There is also a third work by He and Lin \cite{Guansheng-2016}, in which a numerical computation of the deflection angle was made, with agreement with the result of Accioly and Ragusa. Our computation also is consistent with the results of Ref.\cite{Accioly:2002ck}.  It can also be checked conserving the $\mathcal{O}(m^2)$ terms  of the expression \eqref{ugamma} and applying the method proposed in Ref.\cite{Arakida:2011ty} to compute the bending angle.
For a final comment, let us note that by doing the identification $v^2\leftrightarrow 1-\frac{\omega_e^2}{\omega^2_\infty}$ we can use expression \eqref{secondorder} in order to obtain the deflection angle at second order in an homogeneous plasma for light rays,
\begin{eqnarray}
\begin{aligned}
\alpha=&\frac{2m}{b}\bigg(1+\frac{1}{1-{\omega_e^2}/{\omega^2_\infty}}\bigg)\\
&+\frac{3\pi}{4b^2}\bigg(1+\frac{4}{1-{\omega_e^2}/{\omega^2_\infty}}\bigg)m^2+\mathcal{O}(m^3),
\end{aligned}
\end{eqnarray}
which generalizes at second order Eq.\eqref{phs}. However, because of the smallness of the plasma effects, this correction does not appear to be relevant for near-future observations.

\section{Final remarks}
In this work, we have shown how the Gauss-Bonnet theorem can be successfully used to study plasma media in gravitational fields. To use this theorem we have made the following assumptions: the underlying spacetime is static with a timelike Killing vector field $\xi^\alpha$, and, in particular, spherically symmetric and asymptotically flat; it is surrounded by a cold nonmagnetized plasma that is also assumed to be spherically symmetric and at rest with respect to the timelike orbits of $\xi^\alpha$; and the region under study of the light rays is in the weak gravitational regime. Then, using an appropriate Riemannian optical metric that satisfies a Fermat-like variational principle and that is conformal to the induced metric on the spatial slices $\Sigma_t$ of the physical metric (which are orthogonal to $\xi^\alpha$), it follows that the Gibbons-Werner method can be applied. 

In this way, we obtain an invariant and geometrical expression for the deflection angle in terms of geometrical and topological quantities even when in the physical spacetime the light rays do not follow in general null geodesics. Moreover, by using a correspondence between the motion of a massive particle and the dynamics of light rays in a homogeneous plasma, we have successfully applied the Gibbons-Werner method to the study of the deflection angle of massive particles. In particular, we have shown several applications for the case of a homogeneous plasma and some nonhomogeneous profiles. In the last cases, we have only computed the lower-order correction due to the plasma. For a more complete treatment, we should write the equation for the trajectory $\gamma_p$ in a more precise way and integrate in way similar to that for the second order computation of the deflection angle for massive particles. 

The observational relevance of the influence of the plasma in the bending angle and in the associate quantities has been analyzed by different authors and for several astrophysical situations\cite{BisnovatyiKogan:2010ar,Schulze-Koops:2017tkc, BisnovatyiKogan:2008yg, Er:2013efa, 2018MNRAS475867E}. We would like to mention here that the plasma frequency $f=\omega_e/2\pi$ usually takes values from few kHz to 100MHz\cite{Schulze-Koops:2017tkc}. Even when on the surface of the Earth we are limited by the ionosphere to observe only frequencies above 10MHz, there exists radioastronomy projects that consider the idea of putting in orbit 50 or more nanosatellites with low-frequency antennas with a frequency sensitivity in the range of 0.1-10MHz\cite{Budianu-2015,Bentum:2016ekl}. As shown in Table \ref{T1}, for this range of frequencies the deviation in the position of the images (as determined by the Einstein rings), is not negligible. The Schwarzschild metric case was analyzed in the past for a ratio $\omega_e/\omega_\infty$ of the order of $10^{-3}$\cite{Er:2013efa,BisnovatyiKogan:2010ar}. Here we also present the values for observations in a lower range of frequencies and also for other potential exotic objects. In the last cases, when $q>1$,  the influence of the plasma is smaller than in the Schwarzschild spacetime but still potentially detectable. In particular, we can observe that if the Einstein ring without the presence of the plasma takes a value of the order of $1 \text{arcsec}$, then the difference between the optical and the radio-frequency images vary from micro-arcseconds to milli-arcseconds. These  differences should be detectable in the near future.

On the other hand, the Gauss-Bonnet theorem is useful not only for describing weak gravitational lensing but also lensing effects in the strong regime and providing finite distance corrections. 
Note also that, even when we have applied the Gauss-Bonnet theorem to  static and spherically symmetric gravitational fields with a dispersive medium characterized by a refractive index $n(r,\omega(r))$, it can be also applied to nondispersive fluids. It follows as a consequence that in that case the light rays with tangent vectors $\ell^\alpha$ must be null geodesics of the Gordon metric, and therefore they must satisfy the condition $g_{\alpha\beta}\ell^\alpha\ell^\beta=0$, that is \begin{equation}
dt^2=\frac{n^2(r)}{A(r)}(B(r)dr^2+C(r)d\varphi^2),
\end{equation} 
which implies an optical metric as in \eqref{optical-metric}.

To finalize, let us remark that recently a great interest in the study of plasma environments in gravitational fields produced by rotating sources has arisen\cite{Perlick:2017fio,Abdujabbarov:2016hnw,Chakrabarty:2018ase,Atamurotov:2015nra,Liu:2016eju,Dastan:2016bfy,Sharif:2016znp,2013ApSS.346..513M,2016ApSS.361..226A,2017IJMPD..2650051A}. To deal with such situations, a modification of the Gauss-Bonnet theorem approach \cite{Werner:2012rc,Jusufi:2017lsl,Jusufi:2017vew,Jusufi:2017mav,Jusufi:2017uhh,Jusufi:2018jof,Ono:2017pie} can be used. In future works we will show how to apply these techniques for the study of more general plasma environments and the gravitational lensing of massive particles in rotating and stationary gravitational fields.

\begin{table}
    \caption{Relative change in the position of the images for three different frequency ratios $\omega_e/\omega_\infty$ as determined by \eqref{eq:Thetadel} for the class of metrics discussed in Sec. \ref{subex}. Here we assume that $\mu=\gamma$ with $\mu\neq 0$. Note that the same relative change is valid for the deflection angle and the other optical scalars. }
    \label{T1}
   \begin{ruledtabular}
    \begin{tabular}{|c|c @{\qquad} @{\qquad}|c @{\qquad} @{\qquad}|}
        q & ${\omega_{e}}/{\omega_{\infty}}$ & $\frac{\Delta\Theta_0}{\Theta_0}=\frac{\Theta_{pl}-\Theta_{0}}{\Theta_{0}}$  \\
        \hline
        \hline
         & $10^{-1}$ & $2.5\times 10^{-3}$ \\
         1   & $10^{-2}$ & $2.5\times 10^{-5}$ \\
         & $10^{-3}$ & $2.5\times 10^{-7}$  \\
         \hline
         & $10^{-1}$ & $2.0 \times 10^{-3}$ \\
         1.5   & $10^{-2}$ & $2.0 \times 10^{-5}$ \\
         & $10^{-3}$ & $2.0 \times 10^{-7}$  \\
         \hline
         & $10^{-1}$ & $1.7\times 10^{-3}$ \\
         2  & $10^{-2}$ & $1.7\times 10^{-5}$ \\
         & $10^{-3}$ & $1.7\times 10^{-7}$  \\
    \end{tabular}
   \end{ruledtabular}
\end{table}

\section*{Acknowledgments}
We acknowledge support from CONICET and SeCyT-UNC.

\appendix 
\section{Explicit computation of deflection angle for two different non-homogeneous plasma media}\label{appendix}
For completeness, we compute the deflection angle for two different number density profiles. We also make the assumption that $\frac{\omega_e}{\omega_\infty}\ll 1$.
\subsection{Plasma medium with $N(r)=N_{0}r^{-h}, \ h>0$ in a Schwarzschild background}
Here, we consider a gravitational lens surrounded by an inhomogeneous plasma of which the number density of electrons reads
\begin{equation}
N(r)=N_{0}r^{-h}, \ h>0,
\end{equation}
surrounding the exterior of a spherical mass described by the Schwarzschild metric \eqref{schw-ABC}. It follows that the photon frequency has the same behavior as in the homogeneous case; however, the refractive index changes,
\begin{equation}
n(r)=\sqrt{1-\frac{K_{e}N_{0}}{r^{h}\omega^{2}_{\infty}}\bigg(1-\frac{2m}{r} \bigg)}.
\end{equation}
In this case, the associated optical metric is given by,
\begin{equation}\label{schw-inhomo}
\begin{aligned}
d\sigma^{2}=&\frac{r(\omega^{2}_{\infty}-K_{e}N_{0}r^{-h}+2mN_{0}K_{e}r^{-(1+h)})}{\omega^{2}_{\infty}(r-2m)} \\
&\times \bigg(\frac{dr^{2}}{1-\frac{2m}{r}}+r^{2}d\varphi^{2}\bigg),
\end{aligned}
\end{equation}
with determinant $\mathfrak{g}^{\text{opt}}$,
\begin{equation}
\mathfrak{g}^{\text{opt}}=\frac{(r^{h+1}\omega^{2}_{\infty}-K_eN_{0}r+2m K_eN_{0})^{2}r^{3}}{(r-2m)^{3}{}r^{2h}{}\omega^{4}_{\infty}},
\end{equation}
and for the Gaussian curvature, we get
\begin{widetext}
\begin{equation}
\begin{aligned}
\mathcal{K}=&-\frac{1}{2}\frac{\omega^{2}_{\infty}}{r^{3}(r^{h+1}\omega^{2}_{\infty}-K_eN_{0}r+2K_eN_{0}m)^{3}} \times \{8\omega^{2}_{\infty}m^{3}K_eN_{0}(h-\frac{1}{2})(h+3)r^{2h+1}-12\omega^{2}_{\infty}m^{2}K_eN_{0}(\frac{5}{3}h+h^{2}-\frac{3}{2})r^{2h+2} \\
&+6\omega^{2}_{\infty}m K_eN_{0}(h+\frac{3}{2})(h-\frac{2}{3})r^{2h+3}-r^{2h+4}K_eN_{0}h^{2}\omega^{2}_{\infty}+8m[\frac{1}{2}r^{3h+3}\omega^{4}_{\infty}-\frac{3}{4}m r^{3h+2}\omega^{4}_{\infty}+(m^{3}r^{h}-\frac{3}{2}m^{2}r^{h+1}\\
&+\frac{3}{4}m r^{h+2}-\frac{1}{8}r^{h+3})K^2_eN_{0}^{2}(h-2)]\}.
\end{aligned}
\end{equation}
\end{widetext}

To compare with expressions for the bending angle calculated with other methods, we only take into account linear terms in $m$ and $N_{0}$ discarding terms proportional to $mN_{0}$,
\begin{equation}
\begin{aligned}
\mathcal{K}dS=&\bigg(-\frac{2m}{r^{2}}+\frac{h^{2}r^{-(h+1)}K_eN_{0}}{2\omega^{2}_{\infty}}\bigg)dr d\varphi \\ &+\mathcal{O}(m^{2},N_{0}^{2}, mN_{0}).
\end{aligned}
\end{equation}

Furthermore, we need to compute $k_{g}$ and $\frac{dt}{d\varphi}$ along the curve $C_{R}$ associated with the optical metric \eqref{schw-inhomo}, which gives
\begin{equation}
\kappa_g=\frac{R^{\frac{h-3}{2}}\omega_{\infty}|4K_eN_{0}(h-2)(m-\frac{R}{2})^{2}+6R^{h+1}\omega^{2}_{\infty}(\frac{R}{3}-m)|}{(2R^{h+1}\omega^{2}_{\infty}-K_eN_{0}R+2mK_eN_{0})^{3/2}},
\end{equation}
and
\begin{equation}
\frac{d\sigma}{d\varphi}\bigg|_{C_{R}}=\frac{R}{\omega_{\infty}}\sqrt{\frac{\omega^{2}_{\infty}R-R^{1-h}K_eN_{0}+2R^{-h}K_eN_{0}m}{R-2m}}.
\end{equation}
Hence, we can check that
\begin{equation}
\lim_{R\to\infty} \kappa_{g}\frac{d\sigma}{d\varphi}\bigg|_{C_R}=1.
\end{equation}
Finally, using again the expression \eqref{alpha1} the deflection angle reads,
\begin{equation}\label{alpha-inhomo-1}
\alpha=\frac{4m}{b}-\frac{K_eN_{0}}{\omega^{2}_{\infty}}\frac{\sqrt{\pi} \; \Gamma(\frac{h}{2}+\frac{1}{2})}{b^{h} \; \Gamma(\frac{h}{2})} + \mathcal{O}(m^{2},N_{0}^{2}, mN_{0}).
\end{equation}
The expression \eqref{alpha-inhomo-1} agrees with the formula found by Bisnovatyi-Kogan and Tsupko in \cite{BisnovatyiKogan:2010ar}.

\subsection{Plasma medium with $N(r)=N_0e^{-r/r_0}$ in a Schwarzschild background}
For a last example, let us consider a Schwarzschild metric with a plasma medium with a charge number density profile given by 
\begin{equation}\label{eq:Nexp}
N(r)=N_0e^{-r/r_0}.
\end{equation}
In that case the refractive index is given by,
\begin{equation}
n(r)=\sqrt{1-\frac{K_{e}N_0e^{-r/r_0}}{\omega^{2}_{\infty}}\bigg(1-\frac{2m}{r}\bigg)},
\end{equation}
and the associated optical metric is
\begin{equation}\label{schw-inhomo-2}
d\sigma^{2}=\frac{r[\omega^{2}_{\infty}-\omega_{e}^{2}(r)]+2m\omega_{e}^{2}(r)}{(r-2m)\omega^{2}_{\infty}}  \bigg(\frac{dr^{2}}{1-\frac{2m}{r}}+r^{2}d\varphi^{2}\bigg),
\end{equation}
where
\begin{equation}
\omega_{e}^{2}(r)=K_{e}N_0e^{-r/r_0}.
\end{equation}
The determinant $\mathfrak{g}^{\text{opt}}$ of the optical metric \eqref{schw-inhomo-2} reads,
\begin{equation}\label{gexp}
\mathfrak{g}^{\text{opt}}=\frac{[\omega^2_{\infty} r-\omega^2_e(r)(r-2m)]^2r^3}{(r-2m)^3\omega^4_{\infty}},
\end{equation}
while the Gaussian curvature associated with this metric is given by, 
\begin{widetext}
\begin{equation}\label{Kexp}
\begin{aligned}
\mathcal{K}=&\frac{\omega^2_{\infty}}{2r^2_0 r^3[(\omega^2_{\infty}-\omega_e^2(r))r+2\omega^2_e(r) m]^3}\times \{ r(r-2m)\omega^2_{\infty}\omega^2_e(r)[r^4-(4m+r_0)r^3+(4m^2-mr_0)r^{2} \\
+& 6m{r_0}\left({r_0}+m\right)r-6{
m}^{2}{{r_0}}^{2}]
-r_{0}[\omega_{e}^{4}(r)(r-2m)^{3}(m(2r_{0}+r)-r^{2})+2\omega^{4}_{\infty}m r_{0} r^{2}(2r-3m)]
\};
\end{aligned}
\end{equation}
\end{widetext}
which in the case of $N_0=0$ reduces to the expression for the Gaussian curvature of the optical metric associated to the Schwarzschild.
At linear order in $m$, we find
\begin{equation}
\begin{aligned}
\mathcal{K}dS=&\bigg(-\frac{2m}{r^2}+\frac{\omega^{2}_e(r)[\omega^2_{\infty}(r-r_0)+r_0\omega^2_e(r)]}{2r^2_0(\omega^2_{\infty}-\omega^2_e(r))^{2}}\bigg) dr d\varphi\\
&+\mathcal{O}(m^2,N_0m).
\end{aligned}
\end{equation}
The geodesic curvature of $C_R$ reads:
\begin{equation}
\kappa_g=\frac{\sqrt{(R-2R_0)(R-2m)^2w^2_e(R)+2\omega^2_{\infty}r_0R(R-3m)}}{2\omega^{-1}_\infty r_0R^{3/2}[(\omega^2_{\infty}-\omega^2_e(R))R+2mw^2_e(R)]^{3/2}}.
\end{equation}
For this metric we also have
\begin{equation}
\frac{d\sigma}{d\varphi}\bigg|_{C_R}=\frac{R}{\omega_{\infty}}\bigg\{\frac{[\omega^2_{\infty}-\omega^2_e(R)]R+2m\omega^2_e(R)}{R-2m}\bigg\}^{1/2},
\end{equation}
and therefore we check again that 
\begin{equation}
\lim_{R\to\infty} \kappa_{g}\frac{d\sigma}{d\varphi}\bigg|_{C_R}=1.
\end{equation}
Finally, the deflection angle follows
\begin{equation}
\alpha=\frac{4m}{b}-\frac{bK_eN_0}{r_0\omega^2_{\infty}}K_0(\frac{b}{r_0})+ \mathcal{O}(m^{2},N_{0}^{2}, mN_{0});
\end{equation}
with $K_0$ the modified zero Bessel function of the second kind. A similar expression obtained using another method can be found in Ref.\cite{Er:2013efa}.


\end{document}